# Observation of Polarization Topology in Magneto-Optical BICs

Yaqi Li[1,2,3,†], Mingyu Tong[2,4,†,*], Fujia Chen[1,2,3], Li Zhang[1,2,3], Rui Zhao[1,2,3], Yuze Hu[2,5], Bei Wu[1,2,3], Haiyao Luo[1,2,3], Hongsheng Chen[1,2,3], Yihao Yang[1,2,3,*]

[1]State Key Laboratory of Extreme Photonics and Instrumentation, ZJU-Hangzhou Global Scientific and Technological Innovation Center, Zhejiang University, Hangzhou 310027, China.

[2]The Electromagnetics Academy at Zhejiang University, College of Information Science and Electronic Engineering, Zhejiang University, Hangzhou 310027, China

[3]Zhejiang Key Lab. of Intelligent Electromagnetic Control and Advanced Electronic Integration, Jinhua Institute of Zhejiang University, Zhejiang University, Jinhua 321099, China

[4]New Cornerstone Science Laboratory, Department of Physics, The University of Hong Kong, Pokfulam Road, Hong Kong, China.

[5]College of Science, National University of Defense Technology, Changsha 410073, China.

†These authors contributed equally: Yaqi Li and Mingyu Tong.

*Correspondence to Mingyu Tong (mytong_research@sina.com) or Yihao Yang (yangyihao@zju.edu.cn)

## Abstract

Polarization topology in momentum space is a fundamental characteristic of photonic bound states in the continuum (BICs), where the winding of far-field polarization defines a quantized topological charge. Although magneto-optical (MO) BICs combine polarization topology with magnetically tunable chirality, experimental characterization remains challenging because time-reversal symmetry (TRS) breaking complicates eigenpolarization extraction. Here, we experimentally resolve the momentum-space polarization topology of an MO BIC by reconstructing its two-dimensional far-field eigenpolarization texture. We extract the resonant output polarization using singular value decomposition of the polarization-resolved angle-resolved reflection matrix without invoking reciprocity. The reconstructed polarization azimuth winds counterclockwise by $2\pi$ around the Γ-point singularity, yielding a topological charge of $q = +1$. Moreover, the polarization handedness can be reversed solely by varying the magnetic-bias strength, while the field direction and the measured topological charge remain unchanged. A non-Hermitian coupled-mode model traces this reversal to hybridization between the BIC mode and a leaky mode. This work establishes experimental access to momentum-space polarization topology in nonreciprocal photonics, opening a route to explore topological singularities and their evolution under active TRS breaking.

## Introduction

Bound states in the continuum (BICs) in photonic crystal slabs are associated with momentum-space polarization singularities, whose topology is encoded in the winding of the surrounding far-field polarization and quantified by a corresponding topological charge [1-5]. This polarization charge fundamentally characterizes BIC topology and has been experimentally determined in conventional reciprocal photonic systems [6-11]. Magneto-optical (MO) BICs combine high-$Q$ resonances and polarization topology with magnetically tunable far-field chirality enabled by breaking time-reversal symmetry (TRS), enriching the momentum-space polarization texture [12-18]. Directly resolving this texture is therefore essential for determining the polarization topological charge and tracking its evolution [19-24], particularly under magnetic tuning [13, 15-18].

Experimentally, however, accessing the far-field eigenpolarization texture surrounding an MO BIC is more challenging than in reciprocal systems. Because TRS is broken, the excitation and radiation polarization channels are no longer simply related by reciprocity [25-28], making it difficult to extract the eigenpolarization from scattering measurements. Consequently, existing experiments on MO BICs have mainly relied on handedness-dependent spectral responses to infer their chirality [29,30]. Such measurements reveal chiral-selective coupling, but do not directly resolve the momentum-space eigenpolarization texture around the BIC. The associated polarization topological charge has therefore remained experimentally unresolved.

In this work, we directly reconstruct the momentum-space eigenpolarization texture surrounding an MO BIC from the full complex polarization-resolved angle-resolved reflection matrix. We extract the resonant output polarization using a singular value decomposition (SVD)-based procedure without invoking a reciprocal relation. The reconstructed polarization azimuth winds counterclockwise by $2\pi$ around the Γ-point polarization singularity, yielding a polarization topological charge of $q = +1$. We further demonstrate that varying the magnetic-bias strength can reverse the polarization handedness, while the field direction and the measured topological charge remain unchanged. An effective non-Hermitian coupled-mode model attributes the chirality reversal to modal hybridization between the BIC mode and a leaky mode. These results establish direct experimental access to the momentum-space polarization topology of MO BICs and reveal the interplay between magnetically reconfigurable chirality and preserved polarization topology.

# Results

## Magnetically tunable BIC in an MO PhC slab

We design an MO PhC slab composed of a square lattice of unit cells periodic in both the $x$ and $y$ directions, with a static magnetic bias $B_0$ applied along the $+z$ direction, as shown in Fig. 1a. The inset shows that the unit cell consists of a yttrium iron garnet (YIG) rod sandwiched between two patterned copper plates, with two permanent magnets placed above and below the slab to provide the magnetic bias. Both the upper and lower patterned copper plates contain L-shaped air apertures, which combine with those of neighboring unit cells to form cross-shaped openings in the periodic lattice (see Supporting Information S1 for details). The underlying geometry exhibits in-plane $C_{4v}$ symmetry and mirror symmetry with respect to the slab mid-plane. Fig. 1b shows photographs of the fabricated sample in front and side views. Within the standard ferrimagnetic gyrotropic model [31,32], the YIG rods under a given magnetic bias are described by the gyrotropic relative permeability tensor

$$\boldsymbol{\mu}_r = \begin{pmatrix} \mu_r & i\kappa & 0 \\ -i\kappa & \mu_r & 0 \\ 0 & 0 & 1 \end{pmatrix}. \tag{1}$$

Here, $\mu_r$ denotes the diagonal in-plane relative permeability, and $\kappa$ denotes the MO coupling coefficient. In all numerical simulations, we fix $\mu_r$ at unity and vary $\kappa$ as an independent effective parameter, with $\kappa = 0$ and $\kappa \neq 0$ representing the absence and presence of MO coupling, respectively. The frequency-dispersive permeability of the YIG under the experimental magnetic bias is provided in Supporting Information S3.

We next establish the BIC character of the relevant eigenmode and examine its persistence under MO coupling. Fig. 1c shows the simulated three-dimensional band structure of the MO PhC near the Γ point for $\kappa = 0$, focusing on the band near $\lambda \approx 20.3$ mm. The color map represents the calculated radiative $Q$ factor. The radiative $Q$ factor increases sharply toward the Γ point and reaches a simulated maximum of $6.8 \times 10^7$, confirming a symmetry-protected (SP) BIC at Γ. Fig. 1d shows the in-plane normalized electric-field distribution $E$ of the corresponding Γ-point eigenmode within one unit cell in the absence of MO coupling. The annular field profile exhibits a monopole-like character, and we denote this mode by $|m\rangle$. With the introduction of MO coupling, the Γ-point eigenmode retains a similar annular field profile, remaining predominantly monopole-like. Fig. 1e further compares the $x$-$z$ cross-sectional $E$ distributions at the Γ point and at $(k_x, k_y) = (0.5, 0.5)\ \pi/a$, including the air regions above and below the slab, for $\kappa = 0$ and a representative nonzero value $\kappa = 0.3$. In both cases, the eigenmode remains confined within the slab with negligible radiation, whereas the off-Γ eigenmode on the same band couples to radiative channels. These field distributions demonstrate that the SP BIC persists after MO coupling is introduced.

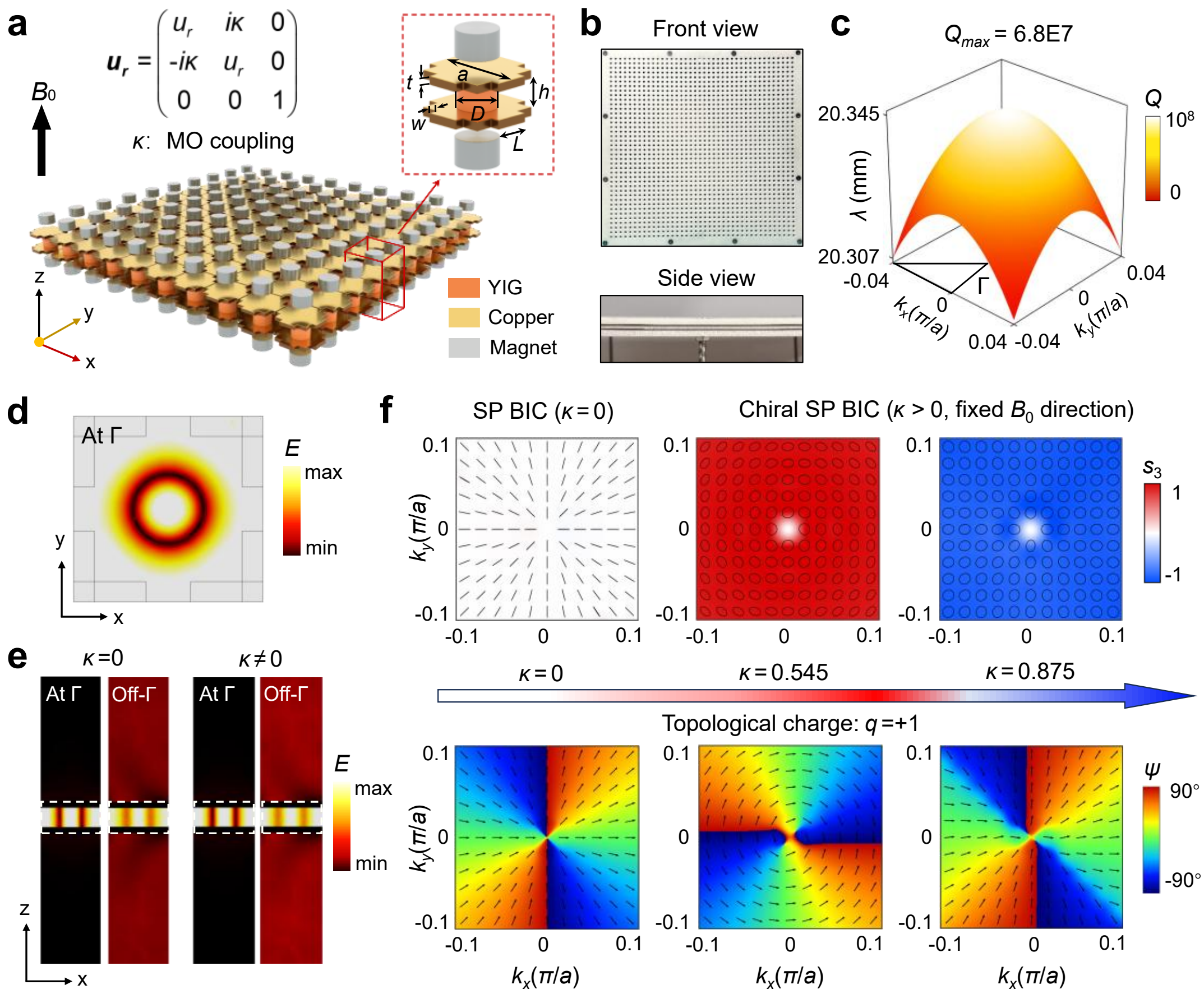

**Fig. 1 Magnetically tunable BIC in an MO PhC slab. a** Schematic of the MO PhC slab biased by a static magnetic field $B_0$ along the $+z$ direction. The left inset shows the gyrotropic relative permeability tensor of the YIG, where $\kappa$ denotes the MO coupling coefficient. The right inset shows the geometry of the unit cell, with $a = 11$ mm, $D = 4.5$

mm, $h$ = 3 mm, $t$ = 1 mm, $w$ = 1.2 mm, and $L$ = 2.6 mm. **b** Photographs of the fabricated sample in front and side views. **c** Simulated three-dimensional band structure of the MO PhC near the Γ point for $\kappa = 0$. The color map denotes the radiative $Q$ factor, which reaches $Q_{max} = 6.8 \times 10^7$ at the Γ point, indicating an SP BIC. **d** Simulated in-plane normalized electric-field distribution $E$ of the Γ-point monopole-like mode $|m\rangle$ within one unit cell in the absence of MO coupling. **e** Simulated $x$-$z$ cross-sectional $E$ distribution at the Γ point and at the off-Γ wavevector $(k_x, k_y) = (0.5, 0.5)\ \pi/a$, for $\kappa = 0$ and $\kappa = 0.3$. The Γ-point eigenmode remains confined in the slab, whereas the off-Γ mode couples to radiative channels. **f** Calculated far-field eigenpolarization textures around the Γ-point BIC for $\kappa$ = 0, 0.545, and 0.875 (from left to right). The upper panels show the $s_3$ distributions overlaid with polarization ellipses, and the lower panels show $\psi$. As $\kappa$ increases with the magnetic-field direction fixed, the surrounding eigenpolarization evolves from nearly linear to right-handed and then left-handed, while $\psi$ retains a counterclockwise $2\pi$ winding and the topological charge remains $q = +1$.

Fig. 1f displays the calculated far-field eigenpolarization textures around the Γ-point BIC for $\kappa$ = 0, 0.545, and 0.875. For a given in-plane wavevector $\boldsymbol{k}_{\|} = (k_x, k_y)$, the complex far-field Jones vector $\boldsymbol{a}(\boldsymbol{k}_{\|}) = [a_x(\boldsymbol{k}_{\|}), a_y(\boldsymbol{k}_{\|})]^{\mathrm{T}}$ is extracted from the simulated eigenmode fields via the projection procedure described in Supporting Information S4. Because the far-field radiation vanishes at the BIC, the chiral and topological characteristics of this singularity are encoded in the far-field eigenpolarization of neighboring radiative eigenstates. Here, we characterize the local polarization handedness using the normalized third Stokes parameter $s_3(\boldsymbol{k}_{\|})$, defined as

$$s_3(\boldsymbol{k}_{\|}) = \frac{S_3}{S_0} = \frac{2\,\mathrm{Im}[a_x^*(\boldsymbol{k}_{\|})a_y(\boldsymbol{k}_{\|})]}{|a_x(\boldsymbol{k}_{\|})|^2 + |a_y(\boldsymbol{k}_{\|})|^2}, \tag{2}$$

where $S_0$ represents the total far-field intensity, while $S_3$ is the third Stokes parameter characterizing the circular-polarization imbalance and handedness. Under the circular-polarization convention defined in Supporting Information S4, $s_3$ = +1 and $s_3$ = -1 correspond to right- and left-handed circular polarization, respectively, while $s_3$ = 0 indicates linear polarization. The polarization azimuth $\psi(\boldsymbol{k}_{\|})$ specifies the orientation of the major axis of the polarization ellipse with respect to the $x$-axis. The winding of $\psi(\boldsymbol{k}_{\|})$ around the momentum-space singularity defines the topological charge evaluated along a counterclockwise closed loop $L$ enclosing the singularity

$$q = \frac{1}{2\pi}\oint_L d\boldsymbol{k}_{\|} \cdot \nabla_{\boldsymbol{k}_{\|}} \psi\left(\boldsymbol{k}_{\|}\right). \tag{3}$$

In the absence of MO coupling ($\kappa = 0$), the radiative states around the SP BIC are nearly linearly polarized with $s_3 \approx 0$, and their polarization orientations form a vortex around the Γ point. At $\kappa$ = 0.545, $s_3$ is predominantly positive and nearly uniform in the near-Γ region ($|k_x|, |k_y| \leq 0.1\ \pi/a$), indicating a right-handed chiral far-field texture. As $\kappa$ increases to 0.875, $s_3$ reverses sign and becomes predominantly negative, corresponding to a left-handed texture. The sign change in $s_3$ indicates a reversal of far-field handedness as the MO coupling strength increases while the magnetic-field direction is fixed. Unlike structurally engineered chiral-BIC platforms [33-37], the chirality here is induced magnetically through TRS breaking without introducing geometric asymmetry. The polarization azimuth $\psi$ retains a counterclockwise $2\pi$ winding around the Γ-point singularity for all three values of $\kappa$, corresponding to a topological charge $q$ = +1 in each case. Extended calculations over the experimentally relevant momentum-space region ($|k_x|, |k_y| \leq 0.4\ \pi/a$) are provided in Supporting Information S5.

## Experimental reconstruction of the eigenpolarization texture of the MO BIC

To experimentally reconstruct the far-field eigenpolarization texture of the MO BIC, we performed polarization-resolved angle-resolved reflection measurements. The fabricated MO PhC sample was assembled in the multilayer configuration shown in Fig. 2a (see Supporting Information S1 for fabrication and assembly details). The same sample was measured under two magnetic-bias strengths, corresponding to the weaker-bias condition in Fig. 2 and the stronger-bias condition in Fig. 3, with the field direction fixed. Fig. 2b illustrates the polarization-resolved angle-resolved reflection measurement setup, consisting of a vector network analyzer (VNA, Ceyear 3672B), a pair of high-collimation-lens antennas (HD-140LHA250+SZJ) serving as the transmitter and receiver, and a motorized rotation stage. In this experiment, the incident angle $\theta$ was varied from -20° to -6° and from 6° to 20°, while the azimuthal angle $\varphi$ was scanned from 0° to 360°. At each accessible momentum point, the four complex reflection coefficients $R_{ss}$, $R_{sp}$, $R_{ps}$, and $R_{pp}$ were measured, as detailed in Supporting Information S2.

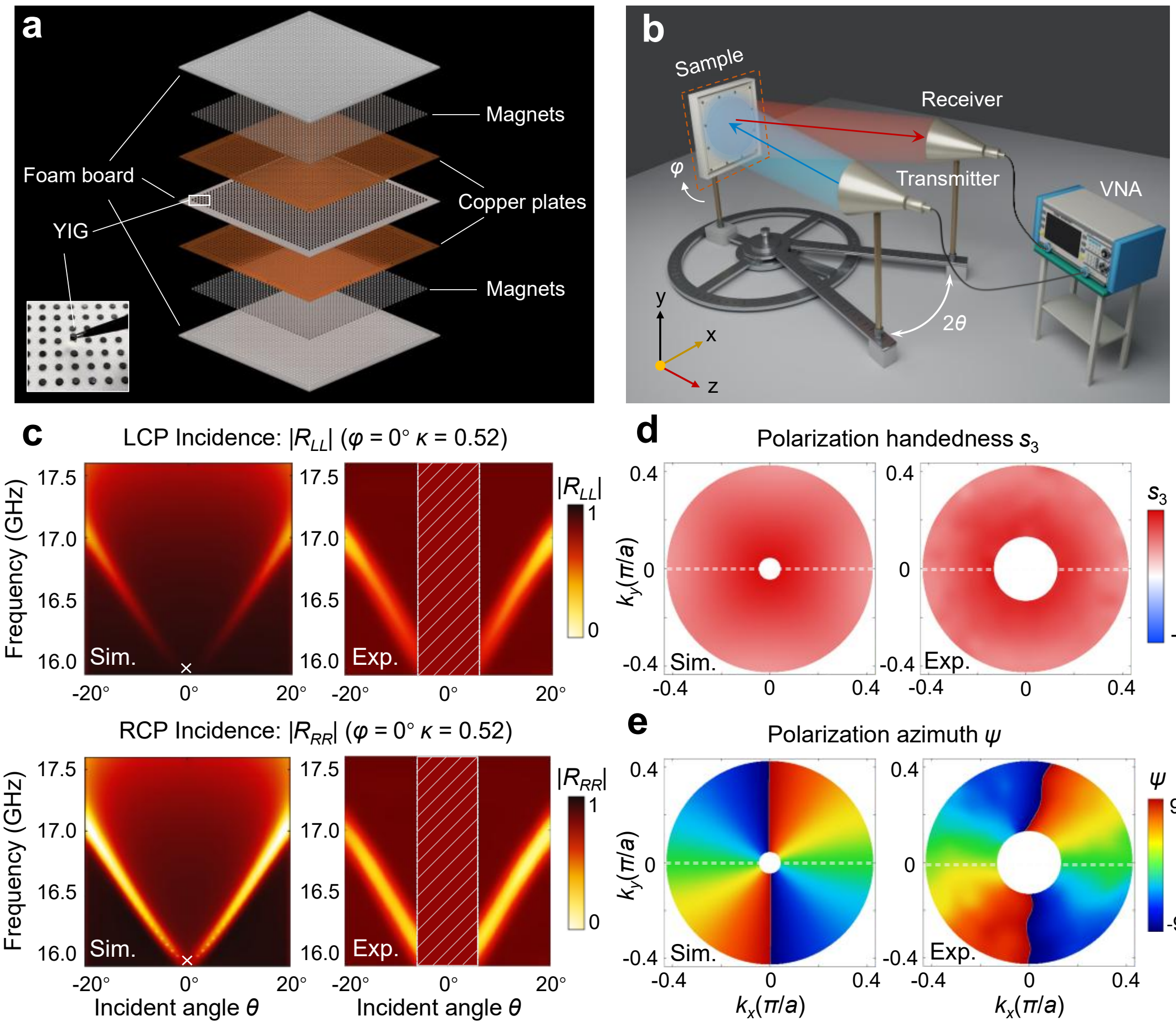

**Fig. 2 Experimental reconstruction of the eigenpolarization texture of the MO BIC. a** Multilayer configuration of the fabricated MO PhC sample. The inset shows a photograph of the YIG rods embedded in the foam board. **b** Schematic of the polarization-resolved angle-resolved reflection measurement setup, consisting of a VNA, the MO PhC sample, a pair of linearly polarized lens antennas, and a motorized rotation platform. **c** Simulated (left column) and measured (right column) co-polarized reflection spectra for LCP (upper row) and RCP (lower row) incidence at $\varphi$ = 0° under the weaker-bias condition, represented in the simulation by $\kappa$ = 0.52. The white crosses mark the simulated Γ-point BIC frequencies, and the hatched regions indicate the experimentally inaccessible angular range

$|\theta| < 6°$. **d, e** Reconstructed momentum-space distributions of $s_3$ (**d**) and $\psi$ (**e**) for the eigenstates surrounding the MO BIC under the same weaker-bias condition. The left and right columns show the simulated and measured results, respectively. White central regions denote momentum-space points excluded from the reconstruction because of weak radiation or experimental inaccessibility. The horizontal white dashed lines in (**d**) and (**e**) indicate the $\varphi = 0°$ ($k_y = 0$) cut corresponding to the spectra in (**c**).

Under the weaker-bias condition, Fig. 2c compares the simulated and measured co-polarized reflection spectra in the circular polarization basis at $\varphi = 0°$. This experimental condition is represented in the simulations by $\kappa = 0.52$, which reproduces the measured response well. The complex circular-basis reflection coefficients are denoted by $R_{ij}$, and their amplitudes by $|R_{ij}|$, where the first and second subscripts refer to the handedness of the reflected and incident waves, with $i, j \in \{L, R\}$. Here, the upper and lower rows show the co-polarized reflection amplitudes $|R_{LL}|$ and $|R_{RR}|$, respectively, while the left and right columns correspond to simulation and experiment. The transformation of the measured linear-basis reflection matrix to the circular-polarization basis is described in Supporting Information S6.

In the simulated spectra, resonance branches appear at finite incident angles but fade toward normal incidence as the radiative coupling approaches zero, consistent with the presence of the SP BIC at the Γ point. The white crosses mark the simulated Γ-point BIC frequencies. Within the accessible angular range, the measured resonance branches agree well with the simulated results. Under the weaker-bias condition, both the simulated and measured spectra exhibit a pronounced resonance in the right-handed circular polarization (RCP) channel $|R_{RR}|$ and a much weaker response in the left-handed circular polarization (LCP) channel $|R_{LL}|$. This handedness-dependent contrast indicates preferential coupling between neighboring radiative eigenstates and the RCP radiation channel. In the experimental spectra, the hatched regions denote the inaccessible angular range $|\theta| < 6°$ due to the limitations of the measurement setup.

We applied SVD to the full four-channel reflection matrix [38-42] to extract the resonant output polarization, as detailed in Supporting Information S7. The reconstructed momentum-space distributions of $s_3$ and $\psi$ are shown in Figs. 2d and 2e, with the left and right columns presenting the simulated and experimental results. The simulated distributions were calculated at the same $\kappa = 0.52$ used for the reflection spectra in Fig. 2c. Both simulation and experiment reveal a predominantly positive $s_3$ over the accessible momentum-space region surrounding the BIC, indicating a right-handed chiral far-field texture. Meanwhile, $\psi$ exhibits a vortex-like distribution with a counterclockwise $2\pi$ winding around the polarization singularity. Despite the inaccessible central region and moderate experimental nonuniformity, the experimental results reproduce the key features predicted by the simulation, directly confirming the chiral far-field eigenpolarization and yielding a topological charge of $q = +1$.

## Chirality reversal of the MO BIC with preserved topological charge

To illustrate how increasing MO coupling with the magnetic-field direction fixed drives the evolution of the far-field chirality, we track the momentum-space eigenpolarization texture as $\kappa$ increases. Figs. 3a and 3b show the simulated momentum-space distributions of $s_3$ and $\psi$ around the Γ-point BIC. As $\kappa$ increases, the surrounding eigenpolarization evolves from nearly linear polarization to a predominantly right-handed chiral texture and then reverses to a predominantly left-handed chiral texture after crossing the chirality-reversal threshold. Fig. 3c quantifies this evolution by plotting the extracted $s_3$ at $(k_x, k_y) = (0, 0.02)\ \pi/a$ as a function of $\kappa$, revealing a clear sign reversal. Meanwhile, $\psi$ maintains a counterclockwise $2\pi$ winding around the central singularity, and the corresponding topological charge remains $q = +1$ throughout the investigated range $0 \leq \kappa \leq 0.9$, as shown in Fig. 3d. The red and blue triangular markers denote the weaker- and stronger-bias conditions, corresponding to $\kappa = 0.52$ and 0.74, respectively. In Fig. 3d, the topological charges indicated by the triangular markers are determined directly from the experimentally reconstructed $\psi$ distributions, with both bias conditions yielding $q = +1$.

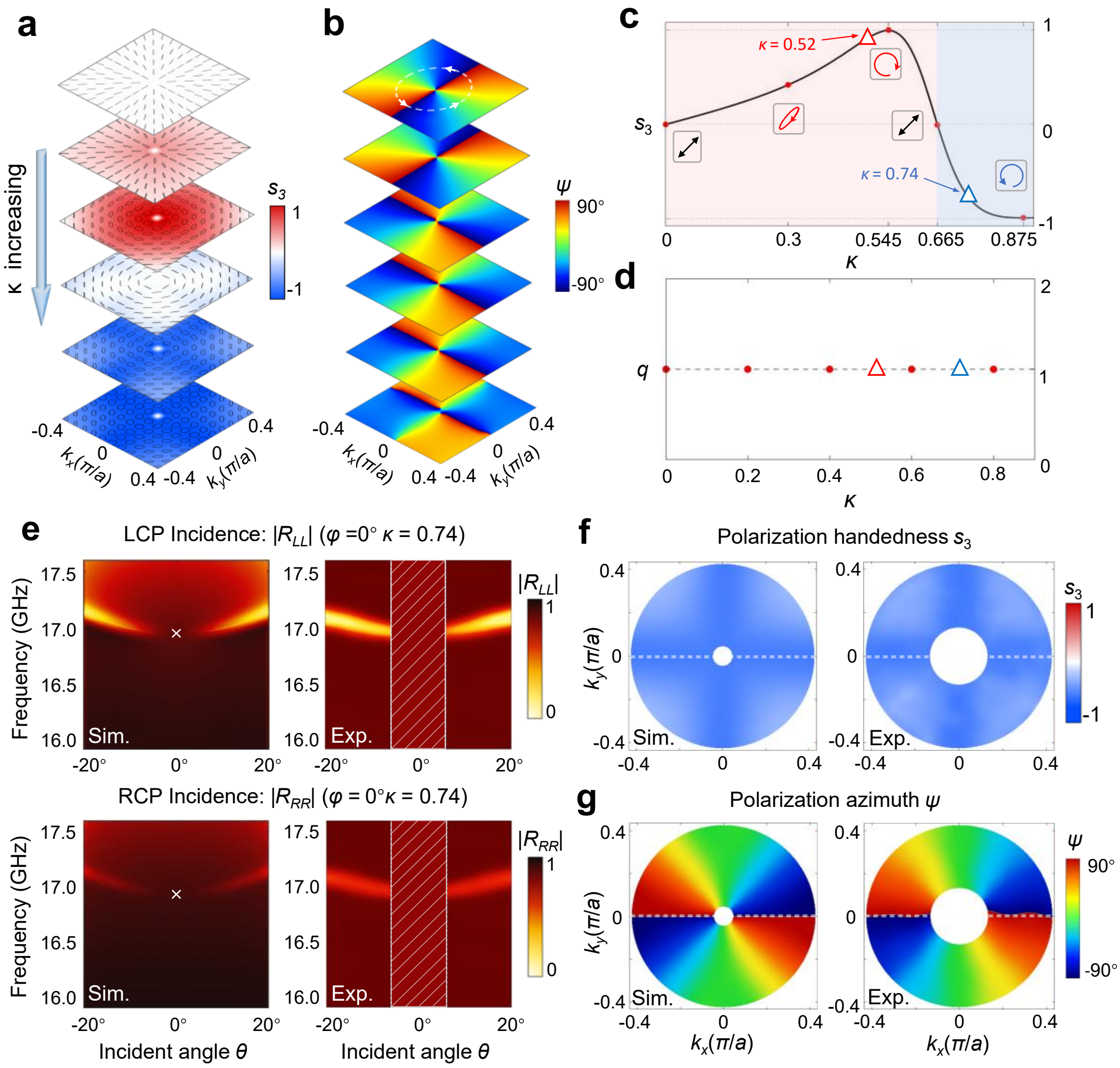

**Fig. 3 Chirality reversal of the MO BIC with preserved topological charge. a, b** Simulated evolution of the momentum-space distributions of $s_3$ (**a**) and $\psi$ (**b**) around the MO BIC as $\kappa$ increases. The background color in (**a**) denotes $s_3$, and the overlaid polarization ellipses represent the local far-field eigenpolarization. The white dashed loop in (**b**) denotes the counterclockwise closed path used to calculate the topological charge in (**d**). **c** Extracted $s_3$ at $(k_x, k_y) = (0, 0.02)\ \pi/a$ as a function of $\kappa$, showing the reversal from right- to left-handed far-field polarization. The light red and light blue backgrounds indicate right-handed and left-handed chirality regions, respectively. The polarization symbols mark representative polarization states at the red solid points on the chirality-evolution curve. **d** Topological charge calculated from the winding of $\psi$ around the singularity as a function of $\kappa$. The red and blue triangular markers in (**c**) and (**d**) mark the weaker- and stronger-bias conditions, corresponding to the simulation parameters $\kappa = 0.52$ and 0.74, respectively. **e** Simulated and measured co-polarized reflection spectra under LCP and RCP incidence at $\varphi = 0°$ under the stronger-bias condition. **f, g** Simulated and experimentally reconstructed momentum-space distributions of $s_3$ (**f**) and $\psi$ (**g**) under the stronger-bias condition. In (**e**-**g**), the white crosses, hatched regions, horizontal white dashed lines, and central white regions follow the same convention as defined in Fig. 2.

Experimentally, we symmetrically reduced the separation between the upper and lower magnet arrays and the YIG rods in the same fabricated sample to increase the magnetic-bias strength, while keeping the bias direction unchanged. The stronger-bias experimental condition is represented in the simulations of Figs. 3e-g by $\kappa = 0.74$. In contrast to the RCP-dominated response under the weaker-bias condition, Fig. 3e shows an LCP-dominated resonant response under the stronger-bias condition. This reversed handedness-dependent contrast indicates a reversal in the chiral-selective coupling of the resonant mode. The corresponding simulated and experimental eigenpolarization textures are shown in Figs. 3f and 3g. Both simulation and experiment exhibit predominantly negative $s_3$ over most of the accessible region surrounding the singularity, while $\psi$ retains a counterclockwise $2\pi$ winding.

## Coupled-mode mechanism underlying the chirality reversal of the MO BIC

To clarify the chirality-reversal mechanism under a fixed magnetic-field direction, we construct an effective non-Hermitian coupled-mode model using the monopole-like mode $|m\rangle$ and the MO-split RCP-like dipole mode $|d_R\rangle$ as the coupled-mode theory (CMT) basis [43-46]. The detailed coupled-mode formulation is provided in Supporting Information S8. These basis modes are distinguished from their hybridized $m$-like and $d_R$-like branches discussed below. As illustrated in Fig. 4a, the modal evolution is governed primarily by the effective detuning $\Delta_{eff}$, defined as the real frequency difference between the $m$-like and $d_R$-like branches at the Γ point. For $\Delta_{eff} > 0$, the $m$-like branch lies above the $d_R$-like branch in real frequency and exhibits right-handed far-field chirality. As $\kappa$ approaches the near-resonant region and $\Delta_{eff}$ approaches zero, the $m$-$d_R$ hybridization becomes pronounced. After the two branches pass through the near-resonant region, their frequency ordering is inverted, while continued tuning of $\kappa$ drives the $m$-like branch through the chirality-reversal threshold and toward the opposite handedness. The coupled-mode analysis shows that detuning-controlled $m$-$d_R$ hybridization modifies the relative weights and phases of the $|m\rangle$ and $|d_R\rangle$ basis components, thereby driving the far-field chirality reversal. The corresponding amplitude and phase evolution is analyzed in Supporting Information S8.

Figs. 4b-d compare the full-wave simulations with the CMT for the $m$-like and $d_R$-like branches. Fig. 4b shows the real eigenfrequencies of the two branches at the Γ point as functions of $\kappa$. The two branches approach each other near $\kappa = 0.545$, corresponding to the near-zero-detuning regime, after which their real-frequency ordering is inverted. Fig. 4c shows the corresponding $Q$ factors at the Γ point. The $m$-like branch retains the SP-BIC character at the Γ point after MO coupling is introduced, with its numerically evaluated radiative $Q$ remaining in the range of $10^6$-$10^8$. This allows the magnetic control of the far-field chirality without compromising the ultrahigh-$Q$ character of the SP BIC. By contrast, the $d_R$-like branch remains radiative at the Γ point, with radiative $Q$ factors of approximately $10^2$ or below. The pronounced difference in radiative loss allows the two branches to remain distinguishable even when their real frequencies nearly coincide. Fig. 4d further shows that the CMT reproduces the simulated chirality evolution of the $m$-like branch at $(k_x, k_y) = (0, 0.02)\ \pi/a$, including the zero crossing near $\kappa = 0.665$, while the $d_R$-like branch remains RCP-like throughout

the investigated $\kappa$ range. These agreements support a detuning-controlled hybridization mechanism for the chirality reversal with a fixed magnetic-field direction.

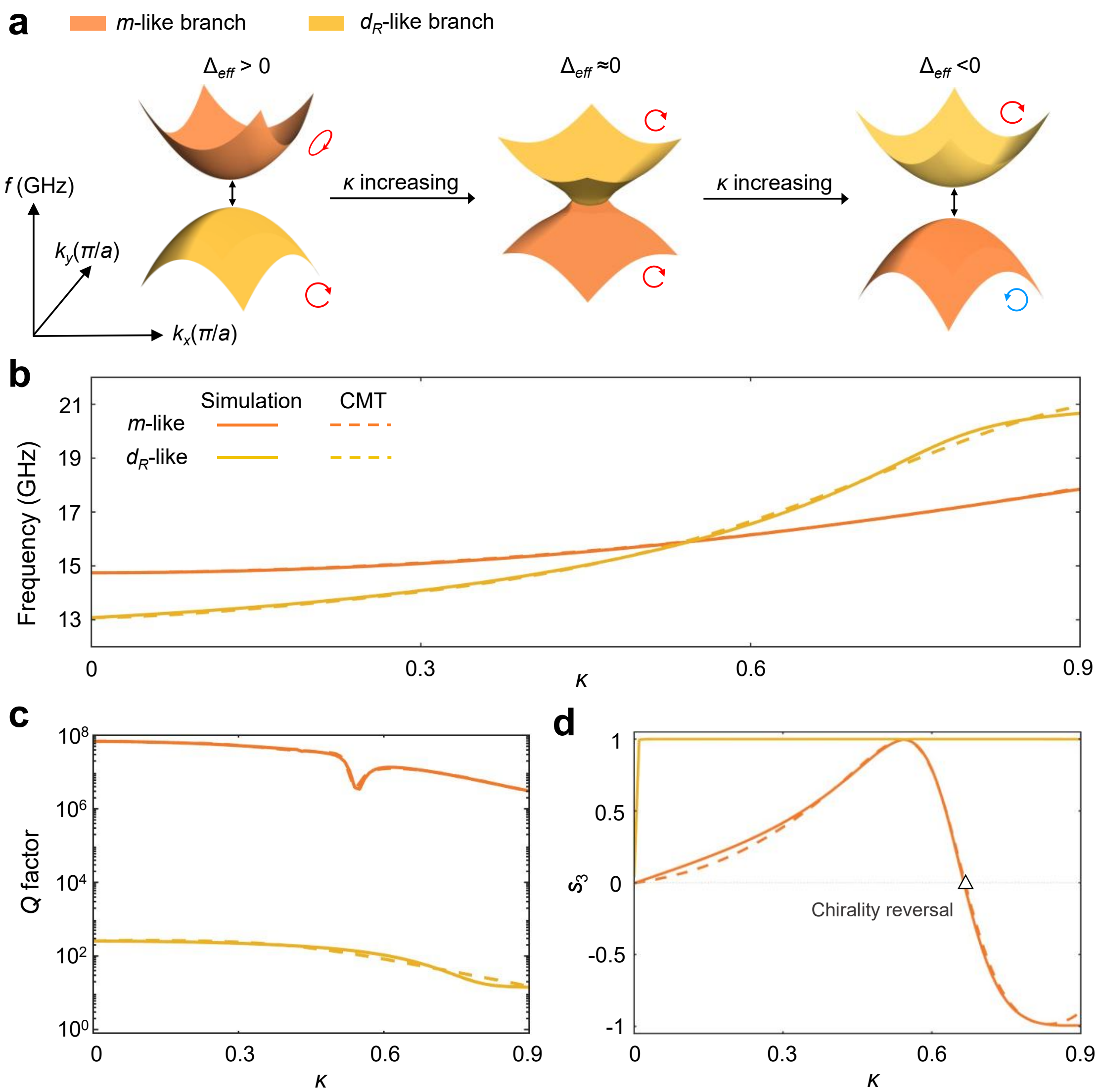

**Fig. 4 Coupled-mode mechanism underlying the chirality reversal of the MO BIC. a** Schematic of the $\kappa$-dependent hybridization between the $m$-like and $d_R$-like branches as the effective detuning $\Delta_{eff}$ evolves from positive (left panel) through near-zero (middle panel) to negative (right panel). Red and blue polarization symbols denote right- and left-handed far-field chirality, respectively, with their shapes schematically representing the polarization ellipticity. **b-d** Comparison between simulations (solid lines) and CMT (dashed lines) for the real eigenfrequencies (**b**) and radiative $Q$ factors (**c**) at the Γ point, and $s_3$ at $(k_x, k_y) = (0, 0.02)\ \pi/a$ (**d**). Orange and yellow curves denote the $m$-like and $d_R$-like branches. The open triangle in (**d**) denotes the zero crossing of $s_3$ for the $m$-like branch.

## Conclusions

In summary, we have experimentally reconstructed the far-field eigenpolarization texture of an MO BIC and determined its momentum-space polarization topological charge. Under two magnetic-bias strengths with the same field direction, the reconstructed eigenpolarization textures exhibit opposite far-field handedness, while the polarization azimuth winds $2\pi$ around the Γ point, yielding $q = +1$ in both cases. Numerical analysis further connects the two experimental states through a continuous chirality reversal as $\kappa$ increases, while the polarization singularity and topological charge remain preserved across the reversal threshold. An effective non-Hermitian coupled-mode model captures this evolution and attributes the chirality reversal to modal hybridization. These results provide direct experimental access to the momentum-space polarization topology of MO BICs and lay the foundation for magnetically controlling topological polarization textures in nonreciprocal photonic systems.

## Methods

**Numerical simulations and far-field polarization reconstruction.** The eigenmode calculations, simulated far-field eigenpolarization textures in Fig. 1, and full-wave reference data in Fig. 4 were obtained using COMSOL Multiphysics. All full-wave simulations employed the reduced gyrotropic model in Eq. (1), with $\mu_r = 1$ and $\kappa$ treated as an independent MO coupling parameter. The complex far-field Jones components were extracted from the simulated eigenmode fields using the projection procedure described in Supporting Information S4. The simulated circular-basis co-polarized reflection spectra in Figs. 2c and 3e were calculated directly in CST Studio Suite in the circular-polarization basis. For the simulated $s_3$ and $\psi$ distributions in Figs. 2d-e and 3f-g, the four-channel complex reflection matrix in the $s/p$ polarization basis was exported from CST and processed using the SVD-based reconstruction procedure described in Supporting Information S7. Representative effective $\kappa$ values for the two experimental bias conditions were selected by jointly matching the simulated band dispersion and reconstructed far-field eigenpolarization texture to the corresponding experimental results.

## Data Availability

The data that support the findings of this study are available from the corresponding author upon reasonable request.

## Authors' Contributions

Y.Y. and M.T. conceived the idea. Y.L. performed the theoretical analysis, developed the sample design and fabrication scheme, and conducted the experimental measurements. M.T. and Y.H. contributed to the theoretical analysis. L.Z., H.L. and B.W. assisted in the sample design. F.C. and R.Z. assisted with the experimental measurements. Y.L., M.T. and Y.Y. drafted and revised the manuscript. Y.Y. and H.C. supervised the project. All authors discussed the results and commented on the manuscript.

## Acknowledgements

This work was supported by the National Key Research and Development Program of the Ministry of Science and Technology of China (Grant Nos. 2022YFA1405200 and 2022YFA1404900), the National Natural Science Foundation of China (Grant Nos. U25D8017, 62305298, and 62305384), the Fundamental Research Funds for the Zhejiang Provincial Universities (Grant No. 226-2025-00231), the Science Challenge Project (Grant No. TZ2025015), and the China National Postdoctoral Program for Innovative Talents (Grant No. BX20230310).

## Conflicts of Interest

The authors declare no conflicts of interest.